\documentclass[aps,pra,amssymb,superscriptaddress,twocolumn]{revtex4-1}
\usepackage[pdftex]{graphicx}
\usepackage[dvipsnames]{xcolor}
\usepackage{bm}
\usepackage{amsmath}
\usepackage{ascmac}
\usepackage{color}
\usepackage{setspace}
\usepackage{amssymb}
\usepackage{latexsym}
\usepackage{mathtools}
\usepackage{lipsum}
\usepackage[normalem]{ulem}
\usepackage[colorlinks,linkcolor=RoyalBlue,citecolor=RoyalBlue,urlcolor=RoyalBlue]{hyperref}
\usepackage{MnSymbol}
\usepackage{comment}

\usepackage{enumerate}
\usepackage{amsthm}
\theoremstyle{plain}
\theoremstyle{definition}

\newcommand{\order}[1]{\mathcal{O}\left(#1\right)}
\newcommand{\mrm}[1]{\mathrm{#1}}
\newcommand{\ket}[1]{\vert #1 \rangle}
\newcommand{\bra}[1]{\langle #1 \vert}
\newcommand{\sket}[1]{\vert #1 \rangle\!\rangle}
\newcommand{\sbra}[1]{\langle\!\langle #1 \vert}
\newcommand{\braket}[1]{\langle #1 \rangle}
\newcommand{\Tr}{\mathrm{Tr}}
\newcommand{\eq}[1]{\begin{equation}\begin{split}#1 \end{split}\end{equation}}
\newcommand{\eqnn}[1]{\begin{equation*}\begin{split}#1 \end{split}\end{equation*}}
\newcommand{\eqs}[1]{\begin{align} #1 \end{align}}

\begin{document}

\title{Exact Current Fluctuations in a Tight-Binding Chain with Dephasing Noise} 

\author{Taiki Ishiyama}
\affiliation{Department of Physics, Institute of Science Tokyo, 2-12-1 Ookayama, Meguro-ku, Tokyo 152-8551, Japan}

\author{Kazuya Fujimoto}
\affiliation{Department of Physics, Institute of Science Tokyo, 2-12-1 Ookayama, Meguro-ku, Tokyo 152-8551, Japan}

\author{Tomohiro Sasamoto}
\affiliation{Department of Physics, Institute of Science Tokyo, 2-12-1 Ookayama, Meguro-ku, Tokyo 152-8551, Japan}

\date{\today}

\begin{abstract}
The full counting statistics (FCS) of current has long provided fundamental insights into nonequilibrium systems.
Recently, the FCS in quantum many-body systems has attracted growing attention,
driven by rapid experimental progress in measuring current fluctuations.
Nevertheless, for diffusive quantum many-body dynamics, the FCS of current has yet to be obtained exactly.
In this Letter, we present the first exact solution for the FCS of current in a diffusive quantum many-body system,
specifically a tight-binding chain with dephasing noise.
By leveraging the system's SU(2) symmetry and a mapping to the one-dimensional Hubbard model,
we derive an exact Fredholm determinant representation for the moment generating function of the time-integrated current.
Our long-time asymptotic analysis shows that the cumulant generating function,
and hence the corresponding large-deviation function, exhibit diffusive scaling for any nonzero dephasing.
We compare our theoretical predictions with experimentally measured current variance and find consistent diffusive scaling.
\end{abstract}

\maketitle 

{\it Introduction.---}
Current is a fundamental quantity in nonequilibrium systems,
and understanding its fluctuations has long been a central challenge in statistical mechanics.
Over the past decades, in classical systems,
the probability distribution of current---rather than only its low-order cumulants---has been intensively investigated~\cite{Lebowitz_1999,Derrida_2007,Seifert_2012}.
This has provided deep insights into universal aspects and unified descriptions of nonequilibrium phenomena,
including the Kardar--Parisi--Zhang universality class~\cite{KPZ,Corwin_2012,Takeuchi_2018} 
and macroscopic fluctuation theory (MFT)~\cite{Bertini_2002,Bertini_2015}.

In recent years, the analysis of the probability distribution of current has been extended to quantum many-body systems~\cite{Schonhammer_2007,Eisler_2013,Tang_2013,Bernard_2013,Znidaric_2014,Znidaric_2014_2,Znidaric_2014_3,Carollo_2017,Yoshimura_2018,Carollo_2018,Moriya_2019,Gamayun_2020,Doyon_2020,Myers_2020,Saenz_2022,Kormos_2022,Nardis_2023,Mcculloch_2023,Bertini_2023,David_2024,David_2024_2,Gamayun_2024,Gopalakrishnan_2024,Gopalakrishnan_2024_2,
Samajdar_2024,Valli_2025,
Costa_2025,Muzzi_2025,Kethepalli_2025,Fujimoto_2026,Fujimoto_2026_2,Yoshimura_2026,Urilyon_2026,Albert_2026, Yadalam_2026},
where it is commonly referred to as full counting statistics (FCS).
This trend has been largely driven by the advent of state-of-the-art experimental platforms such as ultracold atoms and superconducting qubits,
where single-atom- and single-qubit-resolved measurements allow direct access to current fluctuations
in quantum many-body dynamics~\cite{Wei_2022,Wienand_2024,Google,Joshi_2025,Kwon_2026}.
Such experimental advances have, in turn, stimulated theoretical studies of the FCS,
including the development of ballistic MFT~\cite{Doyon_2023_2}, 
which leads to the discovery of novel types of long-range correlations at the ballistic scale~\cite{Doyon_2023}.

Despite these advancements,
an exact solution of the FCS of current in diffusive quantum many-body systems remains lacking.
In diffusive systems,
it is believed that a large-deviation function~\cite{Large,Large_2} 
for current---the nonequilibrium counterpart of a thermodynamic potential---exists,
which implies that all cumulants share the same diffusive scaling,
with the exception of special cases such as integrable systems where convective diffusion occurs~\cite{Nardis_2018,Medenjak_2020,Yoshimura_2025,Yoshimuar_2025_2,Hubner_2025,Doyon_2026,
Fujimoto_2026_2,Yoshimura_2026,Krajnik_2022,Krajnik_2025}.
In classical systems, this fundamental property has been firmly established through exact solutions of the FCS in simple microscopic models, 
such as the symmetric simple exclusion process (SEP)~\cite{Derrida_open,Derrida_current}.
These exact results have played a key role in the development of MFT, a hydrodynamic theory
for describing large deviations of density and current in general diffusive systems.
Recently, extensions of MFT to quantum systems
have been explored in both theoretical and experimental studies~\cite{Wienand_2024, Mcculloch_2023, 
Costa_2025, Albert_2026,
Bauer_2017,Bernard_2019,Bauer_2019,Bernard_2021,Hruza_2023,Alba_2025_2,Russotto_2025,Bernard_2025,Alba_2026}. 
Nevertheless, an exact microscopic derivation of the FCS has 
yet to be achieved for diffusive quantum many-body systems.

\begin{figure}[t]
\begin{center}
\includegraphics[keepaspectratio, width = 6cm]{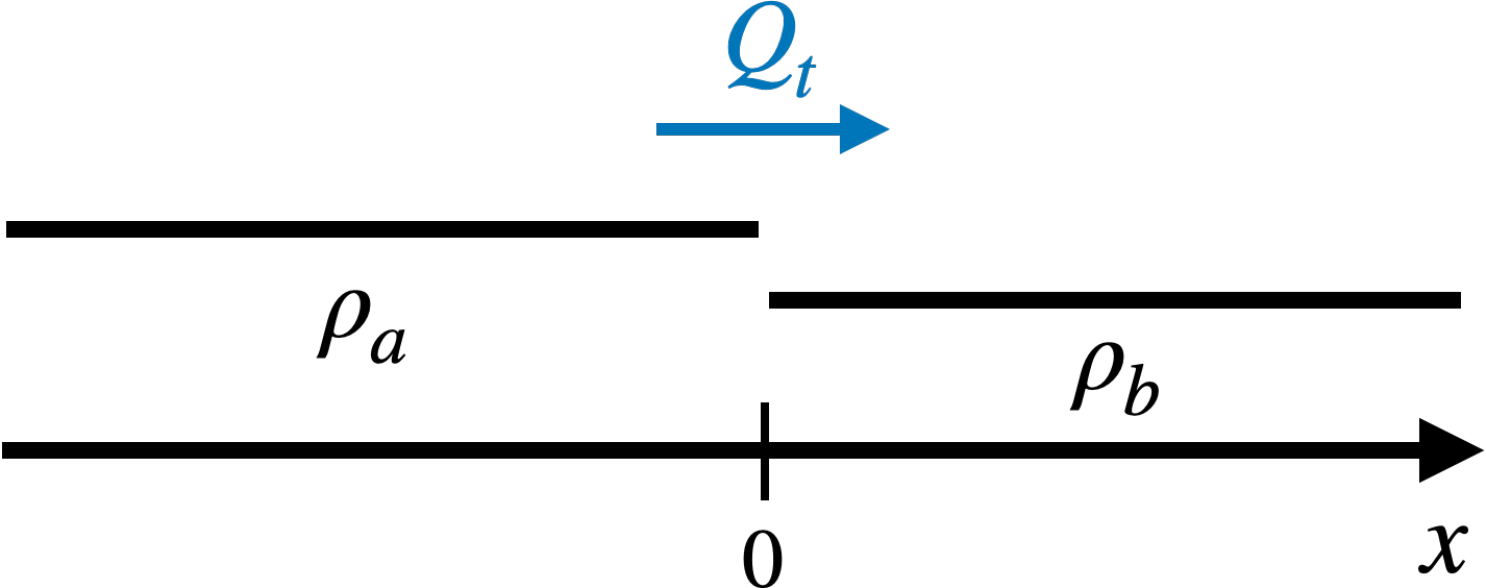}
\caption{
Setup considered in this work.
 The system evolves according to Eq.~\eqref{eq:GKSL} starting from a step initial condition with densities $\rho_a$ and $\rho_b$. 
 We investigate the FCS of the integrated current $Q_t$ across the bond connecting site $0$ and site $1$.
 }
\label{fig:initial}
\end{center}
\end{figure}

In this Letter, we present the first exact solution for the FCS of current in a diffusive quantum many-body system.
We consider a tight-binding chain subject to local dephasing noise as a minimal model for diffusive quantum many-body dynamics.
While the model exhibits ballistic transport in the absence of dephasing~\cite{Antal_1999,Schonhammer_2007,Antal_2008,Moriya_2019,Fujimoto_2025},
it is well established that any nonzero dephasing renders the system diffusive at the level of average and 
variance~\cite{Esposite_2005,Esposite_2005_2,Karevski_2009,Znidaric_2010,Znidaric_2010_2,Znidaric_2011,Znidaric_2014_2,
Medvedyeva_2016,Cao_2019,Turkeshi_2021,Fujimoto_2022,Subhajit_2024,Dhawan_2024,Ishiyama_2025,Alba_2025,Tang_2025,Bhat_2025,Coppola_2025,Tater_2025,Penc_2026,Ray_2026}.
For this model, we derive an exact Fredholm determinant formula for the moment generating function
of the time-integrated current under the step initial condition with two densities (see Fig.~\ref{fig:initial}).
The exact solution is obtained by exploiting the system's $\mathrm{SU}(2)$ symmetry
and a mapping to the one-dimensional Hubbard model with imaginary interaction~\cite{Medvedyeva_exact}.
Through the long-time asymptotic analysis, 
we show that the cumulant generating function, as well as the corresponding large-deviation function,
exhibits diffusive scaling for any nonzero dephasing.
Furthermore, we demonstrate that the cumulant generating function undergoes a crossover from ballistic to diffusive behavior
in the simultaneous limit of weak dephasing and long times.
Finally, we discuss the experimental observability of our results.
In particular, we compare our theoretical prediction with the variance of current observed in a recent experiment~\cite{Kwon_2026},
confirming that both exhibit the same diffusive scaling.

{\it Setup.---}
We consider a tight-binding chain with dephasing noise on an infinite lattice.
The time evolution of the density matrix $\rho(t)$ 
is governed by the Gorini-Kossakowski-Sudarshan-Lindblad (GKSL) equation \cite{Lindblad,Gorini,Breuer},
\begin{equation}
\frac{d \rho}{d t} =\mathcal{L}[\rho]:=-i [H,\rho] +
 \sum_{x \in \mathbb{Z}} [L_x \rho L^\dagger_{x} -\frac{1}{2} \{ L^\dagger_x L_x, \rho \}], \label{eq:GKSL}
\end{equation}
where $\mathcal{L}$ is referred to as the Liouvillian.
The Hamiltonian describes the one-dimensional tight-binding model,
$
H := -\sum_{x\in \mathbb{Z}} (a^\dagger_x a_{x+1} +a^\dagger_{x+1}a_x),
$
where $a_x$ ($a^\dagger_x$) is the annihilation (creation) operator at site $x$.
The dephasing is described by the Lindblad operator $L_x := \sqrt{4\gamma}n_x$, 
with $n_x := a^\dagger_x a_x$ being the number operator and $\gamma$ representing the dephasing strength.
This Lindblad operator can be derived from continuous weak measurements \cite{Breuer, Ashida_dthesis}, 
incoherent light scattering \cite{Pichler_nonequilibrium,Sarker_light}, or noisy on-site potentials \cite{Cehenu_2017}.
In this work, we consider a step initial condition with two densities,
\begin{equation}
\rho_{\mathrm{ini}} := \prod_{x\in \mathbb{Z}} [\rho_x n_x + (1-\rho_x)(1-n_x)],
\label{eq:ini}
\end{equation}
where $\rho_x:=\rho_a$ for $x\leq 0$ and $\rho_x:=\rho_b$ for $x>0$. 
In this state, each site to the left and right of the origin is independently occupied with the probabilities 
$\rho_a$ and $\rho_b$, respectively (see Fig.~\ref{fig:initial}).
Note that this state corresponds to a steady state when $\rho_a = \rho_b$ and
a domain wall state when $(\rho_a,\rho_b)=(1,0)\text{ or }(0,1)$.

Our primary focus is the FCS of the time-integrated current $Q_t$, 
which is fully encoded in the moment generating function (MGF), 
defined as $\braket{e^{\lambda Q_t}} := \sum_{n\in \mathbb{Z}} e^{\lambda n} \mathrm{Pr}[Q_t=n]$.
Here, $Q_t$ represents the net particle flux across the bond connecting site $0$ 
and site $1$ during the time interval $[0, t]$, and $\mathrm{Pr}[Q_t=n]$ denotes the probability that $Q_t$ takes the value $n$.
Formally, due to particle number conservation, 
this is obtained by performing projective measurements of the total particle number in the right subsystem $N_r:= \sum_{x\geq1} n_x$ 
at the initial time $0$ and the final time $t$; 
$Q_t$ is given by the difference between these two measurement outcomes~\cite{Esposito_2009}.
Consequently, $\mathrm{Pr}[Q_t=n]$
is expressed in terms of the projection operator $P_m$ onto the eigenspace of $N_r$ with the eigenvalue $m$:
$
\mrm{Pr}[Q_t=n] = \sum_{m\geq 0} \mathrm{Tr}[P_{m+n}e^{\mathcal{L}t} [P_m\rho_{\mrm{ini}} P_m] ].
$
Summing this probability series yields a compact analytical form for the MGF involving the Liouvillian dynamics:
\begin{equation}
\braket{e^{\lambda Q_t}} = \mathrm{Tr}\left[ e^{\lambda N_r} e^{\mathcal{L}t}
[ e^{-\lambda N_r} \rho_{\mathrm{ini}} ] \right].
\label{eq:mom_genefun}
\end{equation}

{\it Symmetry and integrability framework for $\braket{e^{\lambda Q_t}}$.---}
The foundation of our analysis is the following expression of the MGF,
\eq{
\braket{e^{\lambda Q_t}} = \sum_{n\geq 0}\omega^n q_n(t),\label{eq:omega_dep}
}
where $\omega := \rho_a(1-\rho_b)(e^\lambda-1) + \rho_b(1-\rho_a)(e^{-\lambda}-1)$.
The expansion coefficient reads
\eq{
q_n(t) = \sum_{\substack{y_1<\cdots < y_n \leq 0\\ 0 < x_1<\cdots <x_n}} \prod_{j=1}^n (-1)^{x_j-y_j}
\sbra{ \Phi(\bm{x}) } e^{-it\mathcal{H}} \sket{ \Phi(\bm{y})}.
\label{eq:map_hubbard}
}
Here, $\mathcal{H}$ denotes the one-dimensional Hubbard Hamiltonian with imaginary interaction,
expressed in terms of spin-1/2 fermion operators $c_{x,\sigma}$ and
number operators $n_{x,\sigma}:=c^\dagger_{x,\sigma}c_{x,\sigma}$ ($\sigma=\downarrow,\uparrow$) as
\eq{
\mathcal{H} := &- \sum_{x\in \mathbb{Z},\sigma= \uparrow,\downarrow} (c^\dagger_{x,\sigma} c_{x+1,\sigma} + \mathrm{h.c.} )
\\
&+ 2i\gamma \sum_{x\in \mathbb{Z}} (2n_{x,\uparrow}n_{x,\downarrow} - n_{x,\uparrow} - n_{x,\downarrow}),
\label{eq:hubbard}
}
and $\sket{ \Phi(\bm{x})} := \prod_{j=1}^n c^\dagger_{x_j,\uparrow}c^\dagger_{x_j,\downarrow}\sket{ \Omega}$
denotes the state with $2n$ fermions forming $n$ doublons at sites $\bm{x}$ created from the vacuum $\sket{\Omega}$.
Note that the double-ket notation is introduced to avoid confusion with states in the original Hilbert space.
Crucially, Eq.~\eqref{eq:map_hubbard} implies that the calculation of the MGF is reduced to a finite-particle problem in an exactly solvable system
independent of the parameters $\rho_a$, $\rho_b$, and $\lambda$; namely, the calculation of the propagator for $2n$ particles
governed by the Hubbard Hamiltonian~\cite{Lieb_hubbard,Essler_hubbard}.

The derivation of Eqs.~\eqref{eq:omega_dep} and \eqref{eq:map_hubbard}
proceeds in two steps, relying on the global SU(2) symmetry of the Liouvillian and a mapping to the Hubbard model.
First, we employ the SU(2) symmetry, demonstrating that the MGF depends on $\rho_a$, $\rho_b$, and $\lambda$ solely through the single reduced parameter $\omega$, and that its expansion coefficient reduces to a finite-particle problem.
Define the on-site superoperators $\mathcal{S}_x^\pm$ and $\mathcal{S}_x^z$ acting on the density matrix $\rho$ by
$\mathcal{S}_x^+[\rho]:= a^\dagger_x \rho a_x$,
$\mathcal{S}_x^-[\rho]:= a_x \rho a^\dagger_x$,
and $\mathcal{S}_x^z[\rho]:= (n_x \rho+\rho n_x -\rho)/2$.
These superoperators satisfy the standard SU(2) commutation relations
$[\mathcal{S}_x^+, \mathcal{S}_y^-] = 2\delta_{xy} \mathcal{S}_x^z$ and $[\mathcal{S}_x^z, \mathcal{S}_y^\pm] = \pm \delta_{xy} \mathcal{S}_x^\pm$.
Furthermore, the Liouvillian commutes with the global SU(2) generators: 
$[\mathcal{L}, \sum_{x\in \mathbb{Z}}\mathcal{S}_x^\alpha]=0$ for $\alpha=\pm, z$.
This symmetry implies that the dynamics of infinitely-many particles decomposes into invariant sectors, 
enabling us to reduce the calculation to tractable finite-particle problems.
Specifically, as detailed in End matter, $\braket{e^{\lambda Q_t}}$ takes the form of Eq.~\eqref{eq:omega_dep}
with the coefficient given by the $n$-particle density matrix elements,
\eq{
q_n(t) = \sum_{y_1<\cdots <y_n \leq 0 < x_1<\cdots <x_n} \bra{\bm{x}} e^{\mathcal{L}t }[\ket{\bm{y}}\bra{\bm{y}}] \ket{\bm{x}},
\label{eq:q_n}
}
where we define the Fock basis $\ket{\bm{x}}:= a^\dagger_{x_n}\cdots a^\dagger_{x_1}\ket{0}$.
We remark that a similar parameter dependence of the MGF has been obtained in SEP~\cite{Derrida_current}.
Second, we utilize a mapping to the Hubbard model with imaginary interaction~\cite{Medvedyeva_exact}.
Under a suitable identification of spin-1/2 fermion operators $c_{x,\sigma}$ and a unitary transformation,
the Liouvillian is unitarily equivalent to the Hubbard Hamiltonian, Eq.~\eqref{eq:hubbard}.
As shown in Sec.~\ref{sec:map} of Supplemental Material (SM)~\cite{SM}, applying this mapping to Eq.~\eqref{eq:q_n} yields the expression in Eq.~\eqref{eq:map_hubbard}.

{\it Exact solution of $\braket{e^{\lambda Q_t}}$.---}
We exactly derive the Fredholm determinant representation of the MGF,
\eq{
\braket{e^{\lambda Q_t}} = \det[\hat{I}+\omega \hat{K}^{(\gamma,t)}]_{[0,t]},
\label{eq:freddet}
}
where the kernel $K^{(\gamma,t)}(u,v)$ is defined as
\eq{
K^{(\gamma,t)}(u,v) := \oint \frac{dz}{2\pi i z}\oint \frac{dw}{2\pi i w} \frac{(1+z^2)(1-w^2)}{(z-w)(1+zw)}
\\
\times e^{-u/z + (2t-u)z -2\gamma u } e^{v/w - (2t-v)w -2\gamma v}.
\label{eq:kernel}
}
Here, the counterclockwise contours enclose the origin and are chosen to avoid the poles of the integrand.
Note that the inclusion or exclusion of the poles at $z=w$ and $z=-1/w$ does not affect the result~\cite{RM}.

We outline the derivation of Eq.~\eqref{eq:freddet}.
In our previous work~\cite{Ishiyama_2026},
we derived an integral formula for the multi-particle propagator of the Hubbard model.
Substituting the integral formula into Eq.~\eqref{eq:map_hubbard} yields an exact expression for $q_n(t)$:
\eq{
q_n(t) = \frac{1}{n!}\Big[\prod_{j=1}^n \int_0^t du_j  \Big] \det [K^{(\gamma,t)}(u_j,u_k)]_{j,k=1}^n.
\label{eq:det}
}
Deriving this determinant formula requires techniques analogous to those employed in integrable stochastic interacting systems~\cite{Tracy_asymptotics,Borodin_duality}.
Given the mathematical richness of the derivation, 
we detail the rigorous proof in a separate publication dedicated to the mathematical aspects.
Finally, combining Eqs.~\eqref{eq:omega_dep} and~\eqref{eq:det}, we obtain Eq.~\eqref{eq:freddet}.

\begin{figure}[t]
\begin{center}
\includegraphics[keepaspectratio, width=8.5cm]{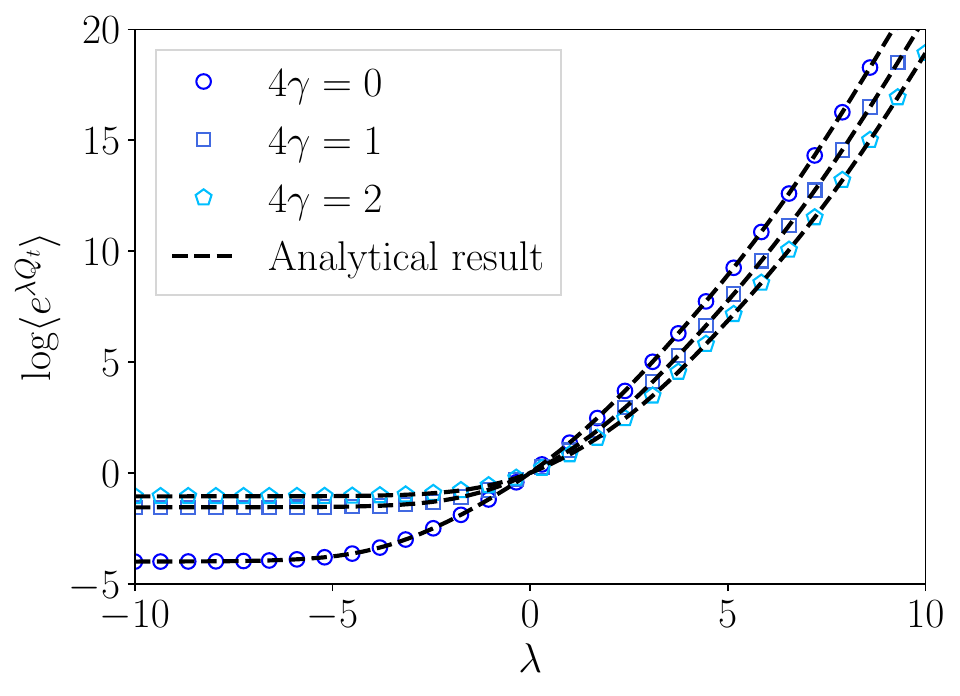}
\caption{
Validation of the exact solution for the MGF at time $t=2$
with the domain wall initial condition, $\rho_a=1$ and $\rho_b=0$.
Symbols show numerical data for a system of $256$ sites
with dephasing rates $4\gamma=0$ (circles), $1$ (squares), and $2$ (pentagons),
while black dashed lines correspond to the Fredholm determinant
in Eq.~\eqref{eq:freddet}, evaluated numerically~\cite{Bornemann}.
}
\label{fig:det}
\end{center}
\end{figure}

We validate the exact solution in Eq.~\eqref{eq:freddet} by comparing it with numerical simulations.
Since the MGF depends on $\rho_a$, $\rho_b$, and $\lambda$ solely through the parameter $\omega$, we focus on the domain wall initial condition ($\rho_a=1$ and $\rho_b=0$) without loss of generality.
As shown in Fig.~\ref{fig:det}, the numerical results exhibit excellent agreement with the analytical predictions.
Details of the numerical method are provided in Sec.~\ref{sec:num} of SM~\cite{SM}.

{\it Long-time behavior of current fluctuations.---}
We investigate the long-time behavior of current fluctuations through the asymptotic analysis of the cumulant generating function (CGF),
defined as $\log \braket{e^{\lambda Q_t}}$.

We first consider the long-time limit $t\to \infty$ with finite dephasing $\gamma>0$.
In this limit, the kernel in Eq.~\eqref{eq:kernel} is of order one only for finite $u$ and $v$, and takes the asymptotic form
\eq{
K^{(\gamma,t)}(u,v) \simeq \frac{e^{-2\gamma (u+v)}}{2\pi (uv)^{1/4}} \frac{\sin [2\sqrt{2t}(\sqrt{u}-\sqrt{v})]}{\sqrt{u}-\sqrt{v}}.
\label{eq:asym_kernel}
}
Substituting this into Eq.~\eqref{eq:freddet}, we obtain the long-time behavior of the CGF for $\gamma>0$,
\eq{
\log \braket{e^{\lambda Q_t}} \simeq \sqrt{\frac{t}{2\gamma}}\int_{-\infty}^{\infty} \frac{dk}{\pi} \log (1+\omega e^{-k^2}).
\label{eq:asym_cumu}
}
The derivations are given in Sec.~\ref{sec:asym} of the Supplemental Material~\cite{SM}.
Crucially, Eq.~\eqref{eq:asym_cumu} implies that all cumulants grow diffusively ($\propto \sqrt{t}$) for any nonzero dephasing.
We note that the large-deviation function $I(q) := - \lim_{t\to\infty}  \log \mathrm{Pr}[Q_t = \sqrt{t}\,q]/\sqrt{t}$
can be obtained from Eq.~\eqref{eq:asym_cumu} via a Legendre transform~\cite{Large_2}.

After an appropriate time rescaling, Eq.~\eqref{eq:asym_cumu} is equivalent to the long-time asymptotic form of the CGF for SEP, 
which was previously derived from a microscopic calculation~\cite{Derrida_current} and MFT~\cite{Mallick_exact}. 
Indeed, perturbative analyses in the strong-dephasing limit~\cite{Cai_algebraic,Bauer_stochastic} 
and MFT arguments at finite dephasing~\cite{Znidaric_2014_2,Carollo_2017} 
have suggested that the long-time behavior of our model coincides with that of SEP. 
Our result exactly establishes this equivalence for any $\gamma>0$, 
thereby providing an exact microscopic validation of MFT for a quantum many-body system.
We also note that a recent study on the quantum SEP established consistency with MFT 
for steady-state currents at the level of individual noise realizations,
under certain assumptions on the scaling behavior of correlation functions~\cite{Albert_2026}.

We next consider the weak-dephasing $\gamma\to 0$ and the long-time $t\to \infty$ limit with {\it fixed} $\gamma t = \order{1}$.
In this limit, the CGF takes the following asymptotic form,
\eq{
\gamma \log \braket{e^{\lambda Q_t}} \simeq \frac{2\gamma t}{\pi } \int_0^1 ds \log (1+\omega e^{-4\gamma t(1-\sqrt{1-s^2})}).
\label{eq:cross_cumu}
}
Details of the derivation are provided in Sec.~\ref{sec:asym} of SM~\cite{SM}.
This formula interpolates between the two transport regimes: 
for $\gamma t \ll 1$, Eq.~\eqref{eq:cross_cumu} reduces to the ballistic result obtained for $\gamma=0$~\cite{Schonhammer_2007}, 
$\log \braket{e^{\lambda Q_t}}  \simeq 2t \log (1+\omega)/\pi $,
whereas for $\gamma t \gg 1$, it recovers the diffusive result, Eq.~\eqref{eq:asym_cumu}. 
Thus, Eq.~\eqref{eq:cross_cumu} elucidates the crossover from ballistic to diffusive current fluctuations induced by dephasing.
We remark that, at the level of average current, 
Ref.~\cite{Cao_2019} has already shown this crossover numerically.

\begin{figure}[t]
\begin{center}
\includegraphics[keepaspectratio, width=8.5cm]{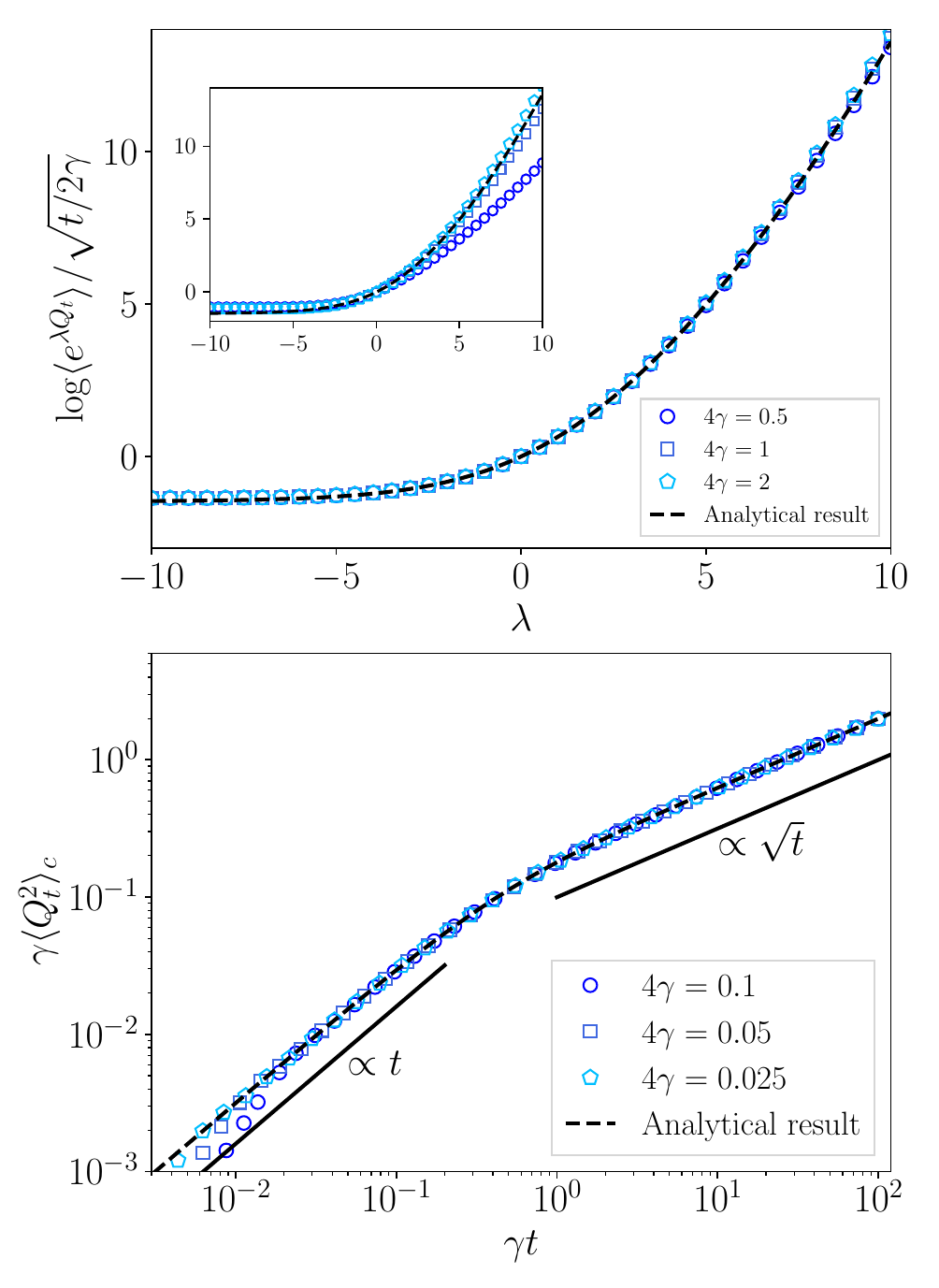}
\caption{
Numerical verification of the asymptotic forms.
(Top panel) Long-time behavior of the CGF under finite dephasing.
Symbols represent numerical results for the domain-wall state
($\rho_a=1$, $\rho_b=0$) for a system of $256$ sites
with dephasing rates $4\gamma = 0.5$ (circles), $1$ (squares), and $2$ (pentagons).
The main panel and the inset display the results for scaled times
$t/2\gamma = 400$ and $t/2\gamma = 20$, respectively.
Dashed lines indicate the asymptotic form in Eq.~\eqref{eq:asym_cumu}.
(Bottom panel) Crossover from ballistic to diffusive fluctuations
for the second cumulant.
Symbols show numerical data for the steady state
($\rho_a=\rho_b=1/2$) for a system of $4096$ sites
with dephasing rates $4\gamma = 0.1$ (circles), $0.05$ (squares),
and $0.025$ (pentagons).
The dashed curve corresponds to the asymptotic form of
$\braket{Q_t^2}_c$ derived from Eq.~\eqref{eq:cross_cumu},
while solid lines serve as guides to the eye,
indicating proportionalities to $t$ and $\sqrt{t}$.
}
\label{fig:long}
\end{center}
\end{figure}

We numerically validate our analytical predictions for the asymptotic behaviors.
Details of the numerical methods are provided in Sec.~\ref{sec:num} of SM~\cite{SM}.
The top panel of Fig.~\ref{fig:long} displays the CGF for the domain-wall state ($\rho_a=1,\rho_b=0$) with various values of $\gamma$.
The results show excellent agreement with the asymptotic formula Eq.~\eqref{eq:asym_cumu},
confirming that current fluctuations exhibit diffusive growth under nonzero dephasing.
In the bottom panel of Fig.~\ref{fig:long}, we focus on the weak dephasing regime.
Due to the numerical difficulty in evaluating the full CGF in this case, 
we analyze the second cumulant $\braket{Q_t^2}_c:=\partial^2_\lambda \log \braket{e^{\lambda Q_t}}|_{\lambda=0}$ 
for the steady state ($\rho_a=\rho_b=1/2$).
The numerical data agree well with the theoretical prediction derived from Eq.~\eqref{eq:cross_cumu},
clearly demonstrating that the current fluctuations exhibit the ballistic-to-diffusive crossover.

{\it Discussion on experimental realizations.---}
We discuss the experimental observability of our theoretical predictions.
In particular, we compare our result with a recent ultracold atom experiment~\cite{Kwon_2026},
which observed diffusive growth of the current variance in the XX spin chain
under a spatiotemporal random potential.
This experimental system effectively reduces to our model in the limit
where the random potential becomes white noise~\cite{Cehenu_2017},
although in the actual experiment the noise exhibits finite spatiotemporal correlations.

For a quantitative comparison, two issues must be addressed.
First, we determine the dephasing strength $\gamma$ corresponding to the experimental setup.
We fix $\gamma$ by matching the experimentally measured diffusion constant, $D=0.91$, 
with that of our model, $D=1/(2\gamma)$.
Second, we account for the difference in initial conditions.
While the experiment employs an alternating initial condition,
in which the initial density has no fluctuations,
our setup corresponds to an initial condition with density fluctuations.
This difference can be addressed within the framework of MFT.
In particular, for systems with particle-hole symmetry,
the CGF for the non-fluctuating initial condition at average density $1/2$
can be related to that of our initial condition with $\rho_a=\rho_b=1/2$~\cite{Derrida_current2}.
Using this relation, one obtains the variance for the alternating initial condition as
$
\braket{Q_t^2}_c \simeq (2\sqrt{2\pi})^{-1}\sqrt{t/(2\gamma)}.
$

\begin{figure}[t]
\begin{center}
\includegraphics[keepaspectratio, width=8.5cm]{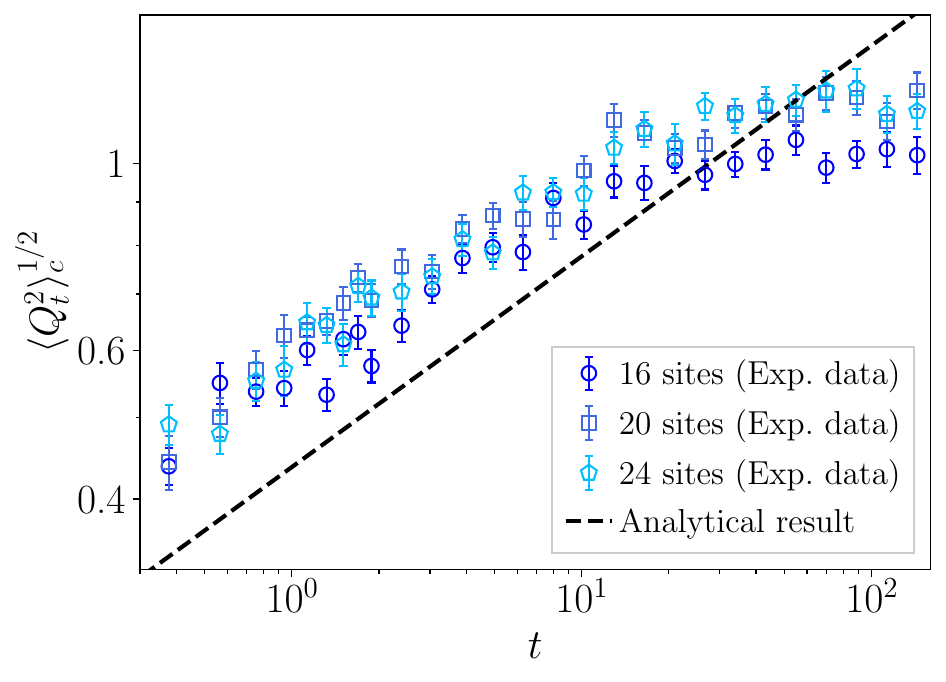}
\caption{
Comparison between our theoretical prediction and the experimental data.
Symbols represent the experimental data from Fig.~4b of Ref.~\cite{Kwon_2026} 
for system sizes of 16 (circles), 20 (squares), and 24 (pentagons) sites.
The dashed line shows the theoretical prediction, $\braket{Q^2}_c \simeq (2\sqrt{2\pi})^{-1}\sqrt{t/(2\gamma)}$, with $1/(2\gamma)=0.91$.
The experimental data are fitted in the range $1 \le t \le 10$ by $Ct^{1/4}$.
The fitted value of $C$ deviates from the theoretical prediction by about $28\%$.
}
\label{fig:ex}
\end{center}
\end{figure}

In Fig.~\ref{fig:ex}, we compare our theoretical prediction with the experimental data.
Both results exhibit diffusive growth, and the prefactors are in reasonable agreement,
albeit with a discrepancy of about $28\%$ between the fitted value and the theoretical prediction.
This discrepancy may be attributed to experimental imperfections, 
including finite correlations in the noise and deviations from the ideal alternating initial condition.
Achieving a more quantitative agreement at the level of the prefactor
may require a refined analysis incorporating such experimental effects.
Nevertheless, the experiment successfully measures the variance of the current,
demonstrating that diffusive current fluctuations are accessible in current experimental setups~\cite{Kwon_2026}.
This suggests that the diffusive scaling of the CGF and the large-deviation function 
could also be observed within the same framework.
Finally, we remark that
observing the ballistic-to-diffusive crossover may require system sizes and times beyond those accessible in current experiments
as seen in the bottom panel of Fig.~\ref{fig:long}.
We expect that such behavior will be observed in future experiments.

{\it Conclusion and future prospects.---}
In this Letter, we studied the FCS of current in a tight-binding chain subject to dephasing noise. 
We have derived an exact expression for the moment generating function under the step initial condition 
by exploiting the SU(2) symmetry and integrability of the system. 
Our asymptotic analysis shows that the cumulant generating function exhibits diffusive growth in the long-time limit $t\to \infty$ 
for any nonzero dephasing $\gamma>0$ and undergoes a crossover from ballistic to diffusive behavior 
in the simultaneous limit of weak dephasing and long times, characterized by $\gamma t \sim 1$. 
Furthermore, we have discussed the experimental observability of our results by comparing our theoretical prediction with a recent experiment~\cite{Kwon_2026}, finding consistent diffusive scaling.

An intriguing future direction is to explore broader classes of nonequilibrium fluctuations in quantum many-body systems
by extending our approach based on symmetry and integrability.
An immediate application would be to a GKSL system~\cite{Ziolkowska_2020} admitting a mapping to the Maassarani model~\cite{Maassarani_1998},
which is an SU(N) generalization of the Hubbard model.
Furthermore, characterizing the ballistic-to-diffusive crossover from a hydrodynamic perspective is an important step
toward establishing a unified description of transport that encompasses both integrable and chaotic regimes~\cite{Caux_2019,Friedman_2020,Bastianello_2020,Durnin_2021,Bastianello_2021,Piqueres_2021,Frederik_2021,Piqueres_2022,Cataldini_2022,Piqueres_2023,Panfil_2023,Lebek_2024,Biagetti_2026,Santra_2026}.

\begin{acknowledgments}
The authors are grateful to Kiryang Kwon and Jae-yoon Choi 
for providing the experimental data used in this work~\cite{Kwon_2026}.
We also thank Fabian Essler, Cristian Giardin\`a, Takashi Imamura, Matteo Mucciconi, and 
Hayate Suda for helpful discussions and comments.
The work of TI has been supported by JST SPRING, Japan Grant Number JPMJSP2180.
The work of KF has been supported by JSPS KAKENHI Grant No. JP23K13029. 
The work of TS has been supported by JSPS KAKENHI Grants (Nos. JP21H04432, JP22H01143, and JP23K22414).
\end{acknowledgments}

\bibliography{ref_revtex}

\begin{thebibliography}{136}
\bibitem [{RM()}]{RM}%
  \BibitemOpen
  \href@noop {} {\ }\bibinfo {note} {After the substitution $w\to1/w$, the
  residues at $z=w$ and $z=-1/w$ change sign. Hence their contributions cancel
  upon performing the $w$-integration.}\BibitemShut {Stop}%
\end{thebibliography}

\section*{End Matter}
{\it Derivation of Eqs.~\eqref{eq:omega_dep} and \eqref{eq:q_n}.---}
We show that the MGF takes the form of Eq.~\eqref{eq:omega_dep} with Eq.~\eqref{eq:q_n}.
In the derivation, we consider a finite lattice $\Lambda_L :=\{x\in \mathbb{Z}\mid -L<x\leq L\}$ with open boundary conditions.
The desired result is obtained by taking the limit $L\to \infty$.
We remark that a similar result for the MGF has been derived for SEP~\cite{Derrida_current}.
While a derivation analogous to SEP is detailed in Sec.~\ref{sec:omega} of SM~\cite{SM}, 
here we present an alternative approach utilizing the global SU(2) symmetry, explicitly.

We first introduce the Hilbert-Schmidt inner product for operators $A$ and $B$ as $(A,B) := \mathrm{Tr}[A^\dagger B]$.
From Eq.~\eqref{eq:mom_genefun}, the MGF can be rewritten as
\eq{
\braket{e^{\lambda Q_t}} = (e^{\lambda N_r}, e^{\mathcal{L}t} [e^{-\lambda N_r} \rho_{\mathrm{ini}}]).
}
We use 
the expansion $e^{\lambda N_r} = \sum_{n=0 }^L \sum_{\bm{x}\in \Lambda^{(n)}_{L,r}} (e^\lambda-1)^n n_{x_1}\cdots n_{x_n}$
and the relation $e^{\mathcal{S}^+}[\ket{\bm{x}}\bra{\bm{x}}] = n_{x_1}\cdots n_{x_n}$
for $\bm{x}\in \Lambda^{n}_L$.
Here we defined $\Lambda_L^n := \underbrace{\Lambda_L \times \cdots \times \Lambda_L}_{n~\text{times}}$,
$\Lambda^{(n)}_{L,r} :=\{\bm{x}\in \Lambda_L^n  \mid  0 < x_1<\cdots< x_n\}$, and 
$\mathcal{S}^{+} := \sum_{x\in \Lambda_L}\mathcal{S}^+_{x}$.
This leads to
\eq{
\braket{e^{\lambda Q_t}} &= \sum_{n=0 }^L (e^\lambda-1)^n (\rho_b/\tilde{\rho}_b e^{\lambda})^L 
\\
&\quad \times \sum_{\bm{x}\in \Lambda^{(n)}_{L,r}} (\ket{\bm{x}}\bra{\bm{x}}, e^{\mathcal{L}t} [\rho_a^{N_l} \tilde{\rho}_b^{N_r} ] ),
\label{eq:expand}
}
where we applied the commutation relation $[\mathcal{L}, \mathcal{S}^+]=0$ and defined 
$N_l := \sum_{-L < x \leq 0} n_x $ and $\tilde{\rho}_b := \frac{e^{-\lambda}\rho_b}{1+\rho_b (e^{-\lambda}-1)}$.

We next use the commutation relation $[\mathcal{L},\sum_{x\in \Lambda_L} \mathcal{S}^z_x]=0$.
Noting that $e^{\nu (\mathcal{S}^z_x+1/2)}[e^{\mu n_x}] = e^{(\nu + \mu)n_x}$, one obtains
\eqnn{
(\ket{\bm{x}}\bra{\bm{x}}, e^{\mathcal{L}t} [\rho_a^{ N_l} \tilde{\rho}_b^{N_r} ] )
= \tilde{\rho}_b^n (\ket{\bm{x}}\bra{\bm{x}}, e^{\mathcal{L}t} [(\rho_a/\tilde{\rho}_b)^{N_l}] ).
}
A calculation analogous to the derivation of Eq.~\eqref{eq:expand} yields
\begin{align}
&(\ket{\bm{x}}\bra{\bm{x}}, e^{\mathcal{L}t} [(\rho_a/\tilde{\rho}_b)^{N_l}] )
=\sum_{m=0}^n (\rho_a/\tilde{\rho}_b-1)^m \notag \\
&\quad \times\sum_{\bm{y}\in \Lambda^{(m)}_{L,l}} \sum_{\substack{\bm{x}' \subseteq \bm{x} \\ |\bm{x}'|=m}}
(\ket{\bm{x}'}\bra{\bm{x}'}, e^{\mathcal{L}t} [\ket{\bm{y}}\bra{\bm{y}}] ),
\end{align}
where $\Lambda^{(m)}_{L,l}:=\{\bm{y}\in \Lambda^m_{L} \mid y_1<\cdots <y_m\leq 0\}$.
Substituting this expression into Eq.~\eqref{eq:expand} and exchanging the order of summation,
we obtain
\eq{
&\braket{e^{\lambda Q_t}} = \sum_{m=0}^L \sum_{n=m}^L
(\rho_a/\tilde{\rho}_b-1)^m \rho_b^n (1-e^{-\lambda})^n \notag \\
&\times(1-\rho_b (1-e^{-\lambda}))^{L-n}
\sum_{\substack{\bm{y}\in \Lambda_{L,l}^{(m)} \\ \bm{x}\in \Lambda^{(n)}_{L,r}}} \sum_{\substack{\bm{x}' \subseteq \bm{x} \\ |\bm{x}'|=m}}
(\ket{\bm{x}'}\bra{\bm{x}'}, e^{\mathcal{L}t} [\ket{\bm{y}}\bra{\bm{y}}] ).
}
In this expression, the summation over $\bm{x}$ can be replaced by $\binom{L-m}{ n-m} \sum_{\bm{x}' \in \Lambda^{(m)}_{L,r}}$.
Using the binomial formula $\sum_{n=m}^L \binom{L-m}{n-m} u^n (1-u)^{L-n} = u^m$
and the identity $(\rho_a/\tilde{\rho}_b-1) \rho_b = \omega$,
we obtain the desired result:
\eq{
\braket{e^{\lambda Q_t}} = \sum_{m=0}^L \omega^m \!\!\!
\sum_{\bm{x}\in \Lambda^{(m)}_{L,r}, \bm{y}\in \Lambda^{(m)}_{L,l}} \!\!\! \bra{\bm{x}}e^{\mathcal{L}t}[\ket{\bm{y}}\bra{\bm{y}}]\ket{\bm{x}}.
\label{eq:omega_dep1}
}

We remark that the derivation of Eq.~\eqref{eq:omega_dep1} is closely related to a duality relation 
that reduces the correlation functions of the infinite many-body system to the density matrix elements of a simpler finite-particle system.
While Ref.~\cite{Medvedyeva_exact} established this duality through direct calculations, it can be understood 
as a consequence of the system's symmetry.
Indeed, utilizing the commutation relation $[\mathcal{L},\mathcal{S}^-]=0$, the correlation function for an initial density matrix $\rho$, defined as
$G_t(\bm{x};\bm{y}) := \mathrm{Tr}[a_{x_1}\cdots a_{x_n} e^{\mathcal{L}t}[\rho] a^\dagger_{y_m}\cdots a^\dagger_{y_1}]$,
can be expressed as
\eq{
G_t(\bm{x};\bm{y})= \bra{\bm{x}} e^{\mathcal{L}t}[e^{(-1)^{n+m}\mathcal{S}^-}[\rho]]\ket{\bm{y}}.
\label{eq:duality}
}
It is worth noting that this duality relation and its derivation based on the symmetry are conceptually 
similar to a Markov duality in classical stochastic processes~\cite{Giardina}.
While the Markov duality is well established, the understanding of dualities in quantum systems remains relatively underdeveloped~\cite{Frassek}.
We leave a systematic investigation of such dualities and their algebraic origins in quantum systems as an important direction for future work.

\widetext
\clearpage

\setcounter{equation}{0}
\setcounter{figure}{0}
\setcounter{section}{0}
\setcounter{table}{0}
\renewcommand{\theequation}{S-\arabic{equation}}
\renewcommand{\thefigure}{S-\arabic{figure}}
\renewcommand{\thetable}{S-\arabic{table}}

\onecolumngrid
\section*{Supplemental Material for ``Exact Current Fluctuations in a Tight-Binding Chain with Dephasing Noise''}

\centerline{Taiki Ishiyama, Kazuya Fujimoto, and Tomohiro Sasamoto}
\vspace{3mm}

\centerline{Department of Physics, Institute of Science Tokyo, 2-12-1 Ookayama, Meguro-ku, Tokyo 152-8551, Japan}
\vspace{5mm}

\par\vskip3mm \hrule\vskip.5mm\hrule \vskip.30cm
This Supplemental Material describes the following:
\begin{itemize}
\item[  ]{(I) Alternative derivation of Eqs.~\eqref{eq:omega_dep} and~\eqref{eq:q_n},}
\item[  ]{(II) Derivation of Eq.~\eqref{eq:map_hubbard},}
\item[  ]{ (III) Asymptotic analysis for the CGF, } 
\item[  ]{ (IV) Numerical scheme for the CGF and the second cumulant.} 
\end{itemize}
\par\vskip1mm \hrule\vskip.5mm\hrule

\vspace{5mm}

\section{Alternative derivation of Eqs.~(\ref{eq:omega_dep}) and (\ref{eq:q_n})}
\label{sec:omega}
In End matter, we derive Eqs.~(\ref{eq:omega_dep}) and (\ref{eq:q_n}) of the main text explicitly using the SU(2) symmetry of the system.
Here, we provide an alternative derivation of these results by following the argument in SEP~\cite{Derrida_current}.
Eqs.~(\ref{eq:omega_dep}) and (\ref{eq:q_n}) of the main text read 
\eq{
\braket{e^{\lambda Q_t}} = \sum_{n\geq 0} q_n(t) \omega^n\label{SM:omega_dependence}
}
and 
\eq{
q_n(t) = \sum_{y_1<\cdots<y_n \leq 0 < x_1<\cdots < x_n} \bra{\bm{x}} e^{\mathcal{L}t }[\ket{\bm{y}}\bra{\bm{y}}]\ket{\bm{x}},
\label{SM:q_n}
}
where the parameter $\omega$ is defined as $\omega= \rho_a(1-\rho_b)(e^{\lambda}-1)+\rho_b(1-\rho_a)(e^{-\lambda}-1)$.
In the derivation, we consider a finite lattice $\Lambda_L :=\{x\in \mathbb{Z}\vert -L<x\leq L\}$
with open boundary conditions. We could obtain the desired result by taking the limit $L\to \infty$.

The important observation is that $\langle Q^n_t \rangle$ is a polynomial of degree $n$ in $\rho_a$ and $\rho_b$.
To avoid complexity in the notation, we demonstrate this for $n=1$ and $n=2$.
From Eq.~\eqref{eq:mom_genefun} in the main text, 
the explicit forms of $\langle Q_t \rangle$ and $\langle Q_t^2 \rangle$ are given by
\begin{align}
&\langle Q_t \rangle = \mathrm{Tr}[N_r e^{\mathcal{L}t}[\rho_{\mathrm{ini}}]] - \mathrm{Tr}[N_r \rho_{\mathrm{ini}}] ,
\\
&\langle Q_t^2 \rangle = \mathrm{Tr}[N_r^2 e^{\mathcal{L}t} [\rho_{\mathrm{ini}}]] -2 \mathrm{Tr}[N_r e^{\mathcal{L}t}[N_r\rho_{\mathrm{ini}}] ]
+\mathrm{Tr}[N_r^2 \rho_{\mathrm{ini}}]
\end{align}
with $N_r := \sum^{L}_{x=1}n_x$. 
Note that the terms in the above equations depend on quantities such as $\mathrm{Tr}[n_x e^{\mathcal{L}t}[\rho_{\mathrm{ini}}]]$, $\mathrm{Tr}[n_x e^{\mathcal{L}t}[n_y\rho_{\mathrm{ini}}]]$, and $\mathrm{Tr}[n_x n_y e^{\mathcal{L}t}[\rho_{\mathrm{ini}}]]$.
The first term, $\mathrm{Tr}[n_x e^{\mathcal{L}t}[\rho_{\mathrm{ini}}]]$, is identified as the diagonal element of the 
correlation function $G_t(x;x)$, where the correlation function is given by
\eq{
G_t(\bm{x};\bm{y}) = \mrm{Tr}[a_{x_1}\cdots a_{x_n}e^{\mathcal{L}t}[\rho] a^\dagger_{y_m}\cdots a^\dagger_{y_1}].
}
Since its initial condition is linear in $\rho_a$ and $\rho_b$ and its time evolution is closed via Eq.~\eqref{eq:duality} in End matter, 
this term is strictly linear in the densities.
Similarly, one finds that $\mathrm{Tr}[n_x e^{\mathcal{L}t}[n_y\rho_{\mathrm{ini}}]]$ and $\mathrm{Tr}[n_x n_y e^{\mathcal{L}t}[\rho_{\mathrm{ini}}]]$ are second-order polynomials in $\rho_a$ and $\rho_b$.
Thus, one sees that the $n$-th moment $\braket{Q_t^n}$ ($n=1,2$) is a polynomial of degree $n$ in $\rho_a$ and $\rho_b$.
We remark that these derivations can be extended to general $\braket{Q_t^n}$.

We next substitute the following expression for $\rho_{\mathrm{ini}}$ into Eq.~\eqref{eq:mom_genefun} in the main text,
\begin{equation}
\rho_{\mathrm{ini}} = \sum_{p,q=0}^{L}\sum_{x_1<\cdots < x_p \leq 0 < x_{p+1} <\cdots < x_{p+q}} 
\rho_a^p (1-\rho_a)^{L-p} \rho_b^{q} (1-\rho_b)^{L-q} \vert x_1,\cdots,x_{p+q}\rangle \langle x_1,\cdots,x_{p+q}\vert. 
\end{equation}
Then, we have
\begin{equation}
\begin{split}
\langle e^{\lambda Q_t}\rangle = &\sum_{p,q=0}^{L} 
\rho_a^p (1-\rho_a)^{L-p} \rho_b^{q} (1-\rho_b)^{L-q}e^{-q\lambda}\\
&\times\underbrace{ \sum_{x_1<\cdots < x_p \leq 0 < x_{p+1} <\cdots < x_{p+q}}  
\mathrm{Tr}[e^{\lambda N_r}e^{\mathcal{L}t}[\vert x_1,\cdots,x_{p+q}\rangle \langle x_1,\cdots,x_{p+q}\vert ] ]}_{R_{p+q}(e^{\lambda})}.
 \end{split}\label{eq:genefun_expansion}
\end{equation}
Since the number of total particles is conserved, $R_{p+q}(e^{\lambda})$ is a polynomial of degree $p+q$ in 
$e^{\lambda}$.
Expanding the above equation in powers of $\rho_a$ and $\rho_b$, we obtain
\begin{equation}
\langle e^{\lambda Q_t}\rangle = \sum_{p,q=0}^{L}\rho_a^{p}\rho_b^{q} e^{-q\lambda} S_{p,q}(e^\lambda),
\end{equation}
where $S_{p,q}(e^{\lambda})$ is also a polynomial of degree $p+q$ in $e^{\lambda}$.
To identify the form of $S_{p,q}(e^\lambda)$, we compare the above equation with the following expansion for the case $\lambda \ll 1$,
\begin{equation}
\langle e^{\lambda Q_t}\rangle = \sum_{n=0}^{\infty} \frac{\lambda^n}{n!} \langle Q^n_t\rangle.
\end{equation}
Noting that $\langle Q_t^n\rangle$ is a polynomial of degree $n$ in $\rho_a$ and $\rho_b$, 
one finds that the lowest order of $S_{p,q}(e^{\lambda})$ in $\lambda$ must be at least $\lambda^{p+q}$.
On the other hand, $S_{p,q}(e^{\lambda})$ can be expressed as
\eq{
S_{p,q}(e^\lambda) = \sum_{n=0}^{p+q} a_{n} (e^\lambda -1)^{n}
}
with some constants $a_n$, ($n=0,\cdots, p+q$), since it is a polynomial of degree $p+q$ in $e^{\lambda}$.
From these two conditions, it must have the form
$S_{p,q}(e^{\lambda})=s_{p,q}(e^{\lambda}-1)^{p+q}$,
where $s_{p,q}$ is a constant which does not depend on $\rho_a$, $\rho_b$, and $\lambda$.
Hence, we have
\begin{equation}
\langle e^{\lambda Q_t}\rangle = \sum_{p,q=0}^{L}s_{p,q} [\rho_a(e^{\lambda}-1)]^p [\rho_b(1-e^{-\lambda})]^{q} 
=: G(\rho_a(e^{\lambda }-1), \rho_b (e^{-\lambda}-1)).
\end{equation}
Thus, one can conclude that $\langle e^{\lambda Q_t}\rangle$ is the function
which only depends on the two reduced variables, $\rho_a(e^{\lambda}-1)$ and $\rho_{b}(e^{-\lambda}-1)$.

Next, we use the particle-hole symmetry of the system.
Define a unitary operator as
\begin{equation*}
U := [a^\dagger_{L}-(-1)^{L} a_{L}][a^\dagger_{L-1}-(-1)^{L-1} a_{L-1}]\cdots [a^\dagger_{-L+1}-(-1)^{-L+1} a_{-L+1} ].
\end{equation*}
This operator yields the particle-hole transformation with the change of sign for odd site fermions:
\begin{equation}
U a_j U^\dagger = (-1)^j a^\dagger_j.
\end{equation}
Our model has the particle-hole symmetry, namely
$Ue^{\mathcal{L}t}[\rho]U^\dagger= e^{\mathcal{L}t}[U\rho U^\dagger]$.
Using this, we have
\begin{align}
G(\rho_a(e^{\lambda }-1), \rho_b (e^{-\lambda}-1)) &= \mathrm{Tr}[U^\dagger Ue^{\lambda N_r} e^{\mathcal{L}t}[e^{-\lambda N_r}\rho_{\mathrm{ini}}]]
\\
&= \mathrm{Tr}[e^{-\lambda N_r}e^{\mathcal{L}t}[e^{\lambda N_r} \rho_{\mrm{ini}}]]|_{\rho_a\to 1-\rho_a,~\rho_b\to 1-\rho_b}
\\
&= G((1-\rho_a)(e^{-\lambda}-1), (1-\rho_b)(e^{\lambda}-1)).\label{eq:particle_hole}
\end{align}
In what follows, we assume $\rho_a\leq \rho_b$ without loss of generality.
We also restrict $\lambda$ to be real for the moment
and define $\tilde{\lambda}$ as
the solution of $\rho_b(e^{-\lambda}-1)=e^{-\tilde{\lambda}}-1$.
Then, we have
\eqs{
&\rho_a(e^\lambda-1)= \frac{\rho_a e^{-\tilde{\lambda}}}{\rho_b+ e^{-\tilde{\lambda}}-1}(e^{\tilde{\lambda}}-1),\label{eq:tilde_rho_a}
\\
&0\leq  \frac{\rho_a e^{-\tilde{\lambda}}}{\rho_b+ e^{-\tilde{\lambda}}-1} \leq 1.
}
By setting $\tilde{\rho}_a:=\rho_a e^{-\tilde{\lambda}}/(\rho_b+ e^{-\tilde{\lambda}}-1)$, we can make the following calculation,
\eqs{
\braket{e^{\lambda Q_t}}
&= G(\rho_a(e^{\lambda}-1), \rho_b(e^{-\lambda}-1))
\\
&=G(\tilde{\rho}_a(e^{\tilde{\lambda}}-1), e^{-\tilde{\lambda}}-1)
\\
&=G((1-\tilde{\rho}_a)(e^{-\tilde{\lambda}}-1), 0) 
\\
&= G(\omega,0).
}
Here, we used Eq.~\eqref{eq:tilde_rho_a} in the second line and Eq.~\eqref{eq:particle_hole} in the third line.
The above is the derivation of the $\omega$-dependence for $\lambda \in \mathbb{R}$.
However,
it can be extended to $\lambda \in \mathbb{C}$ by the identity theorem,
since both $\braket{e^{\lambda Q_t}}$ and $G(\omega,0)$
are holomorphic in $\lambda \in \mathbb{C}$ (at least for the finite lattice $\Lambda_L$). 

Finally, we determine the expansion coefficients $q_n(t)$
when $\langle e^{\lambda Q_t}\rangle$ is expanded in terms of $\omega$:
\begin{equation}
\langle e^{\lambda Q_t}\rangle =\sum_{n\geq 0} q_n(t)\omega^n.
\end{equation} 
Since $q_n(t)$ does not depend on $\rho_a$, $\rho_b$, and $\lambda$, we can consider the case $\rho_b = 0$.
In this case, the above equation becomes
\begin{equation}
\langle e^{\lambda Q_t} \rangle = \sum_{n\geq 0} q_n(t)\rho_a^n (e^{\lambda}-1)^n.\label{eq:limit_1}
\end{equation}
On the other hand, Eq.~\eqref{eq:genefun_expansion} for $\rho_b=0$ becomes
\begin{align}
\langle e^{\lambda Q_t}\rangle  &= \sum_{p=0}^{L}\sum_{x_1<\cdots <x_p \leq 0}  \rho_a^p (1-\rho_a)^{L-p}
 \mathrm{Tr}[e^{\lambda N_r} e^{\mathcal{L}t}[\vert x_1,\cdots,x_p\rangle \langle x_1,\cdots,x_{p}\vert ]] \\
&=\sum_{p=0}^{L}\sum_{x_1<\cdots <x_p \leq 0}\sum_{m=0}^{p}  \rho_a^p (1-\rho_a)^{L-p} e^{\lambda m}
\mrm{Pr}_t(N_r = m|x_1,\cdots,x_p)\label{eq:limit_2}.
\end{align}
Here, we used the identity $\sum_{m=0}^{L}P_m=1 $ for the projectors onto eigenspace of $N_r$ with eigenvalue $m$
in the second line,
and defined the probability that $m$ particles are located to the right of the origin $(x > 0)$ at time $t$
with the initial state $\vert x_1,\cdots,x_p\rangle$ as
\begin{equation}
\mathrm{Pr}_t(N_r = m|x_1,\cdots,x_p):=\mathrm{Tr}[P_m e^{\mathcal{L}t}[\vert x_1,\cdots,x_p\rangle \langle x_1,\cdots,x_{p}\vert]].
\end{equation}
Taking simultaneously the limits $\rho_a\to0$ and $\lambda\to\infty$ at fixed $\rho_a e^{\lambda}$ in Eq.~\eqref{eq:limit_1} and Eq.~\eqref{eq:limit_2},
we have the following expression for $q_n(t)$,
\begin{align}
q_n(t) &= \sum_{x_1< \cdots <x_n \leq 0} \mathrm{Pr}_t(N_r = n|x_1,\cdots,x_n)\\
	 &= \sum_{y_1<\cdots <y_n\leq 0 <x_1<\cdots<x_n}   \bra{\bm{x}} e^{\mathcal{L}t }[\ket{\bm{y}}\bra{\bm{y}}]\ket{\bm{x}}.
\end{align}

\section{Derivation of Eq.~(\ref{eq:map_hubbard})}
\label{sec:map}
We derive Eq.~\eqref{eq:map_hubbard} from Eq.~\eqref{eq:q_n} in the main text.
Equation~\eqref{eq:map_hubbard} reads
\eq{
q_n(t) = \sum_{y_1<\cdots <y_n \leq 0 \leq x_1<\cdots < x_n}
\prod_{j=1}^n (-1)^{x_j-y_j} \sbra{\Phi(\bm{x})} e^{-it\mathcal{H}} \sket{ \Phi(\bm{y})},
\label{SM:map_hubbard}
}
where $\mathcal{H}$ denotes the Hubbard Hamiltonian, expressed in terms of 
the spin-up and spin-down fermion operators $c_{x,\uparrow}$ and $c_{x,\downarrow}$ as
\eq{
\mathcal{H} := - \sum_{x\in \mathbb{Z},\sigma= \uparrow,\downarrow} (c^\dagger_{x,\sigma} c_{x+1,\sigma} + \mrm{H.c.} )
+ 2i\gamma \sum_{x\in \mathbb{Z}} (2n_{x,\uparrow} n_{x,\downarrow} - n_{x,\uparrow} - n_{x,\downarrow}),\label{SM:hubbard_second}
}
and $\sket{\Phi(\bm{x})} := \prod_{j=1}^n c^\dagger_{x_j,\uparrow}c^\dagger_{x_j,\downarrow}\sket{\Omega}$ denotes the state with 
$2n$ fermions forming $n$ doublons at sites $\bm{x}$ on the vacuum $\sket{\Omega}$.

We define the superoperators $c_{x,\uparrow}$ and $c_{x,\downarrow}$ as
\eq{
c_{x,\uparrow} [\rho] := a_x \rho,\quad c_{x,\downarrow} [\rho] := P \rho a^\dagger_x,
}
where $P:=\prod_{x\in\mathbb{Z}} e^{i\pi n_x}$. 
We also introduce the Hilbert-Schmidt inner product for the operator space:
\eq{
(A,B)= \mathrm{Tr}[A^\dagger B].
}
Then the adjoint operators of $c_{x,\uparrow}$ and $c_{x,\downarrow}$ are given by
\eq{
c^\dagger_{x,\uparrow} [\rho] = a^\dagger_x \rho,\quad c^\dagger_{x,\downarrow} [\rho]= P \rho a_x.
} 
These superoperators satisfy the canonical anti-commutation relations,
\eq{
\{c_{x,\sigma}, c_{y,\sigma'}\} = 0, \quad
\{c_{x,\sigma}, c_{y,\sigma'}^\dagger\} = \delta_{x,y}\delta_{\sigma,\sigma'}.
}
Therefore they can be regarded as the creation and annihilation operators of spin-1/2 fermions.
The vacuum state annihilated by all $c_{x,\sigma}$ corresponds to the projector onto the original vacuum:
\eq{
\Omega := \ket{0}\bra{0}.
}

In terms of the above superoperators, the Liouvillian is equivalent to the Hubbard Hamiltonian $\mathcal{H}$:
\eq{
iU\mathcal{L}U^\dagger =- \sum_{x\in \mathbb{Z},\sigma= \uparrow,\downarrow} (c^\dagger_{x,\sigma} c_{x+1,\sigma} + \mrm{H.c.} )
+ 2i\gamma \sum_{x\in \mathbb{Z}} (2n_{x,\uparrow} n_{x,\downarrow} - n_{x,\uparrow} - n_{x,\downarrow}) =\mathcal{H},
}
where we define
\eq{
&U := \prod_{x\in\mathbb{Z}} e^{i\pi n_{2x-1,\downarrow}},
\\
&n_{x,\sigma}:= c^\dagger_{x,\sigma} c_{x,\sigma}.
}
Furthermore, it follows that
\eqs{
q_n(t) &= \sum_{y_1<\cdots <y_n \leq 0 \leq x_1<\cdots < x_n}
\bra{\bm{x}} e^{\mathcal{L}t}[\ket{\bm{y}}\bra{\bm{y}}]\ket{\bm{x}} 
\\
&=\sum_{y_1<\cdots <y_n \leq 0 \leq x_1<\cdots < x_n}
(\ket{\bm{x}}\bra{\bm{x}},e^{\mathcal{L}t} \ket{\bm{y}}\bra{\bm{y}})
\\
&=\sum_{y_1<\cdots <y_n \leq 0 \leq x_1<\cdots < x_n} (
U c^\dagger_{x_n,\uparrow}\cdots c^\dagger_{x_1,\uparrow}
c^\dagger_{x_n,\downarrow}\cdots c^\dagger_{x_1,\downarrow}\Omega,
e^{-it \mathcal{H}}U 
c^\dagger_{y_n,\uparrow}\cdots c^\dagger_{y_1,\uparrow}
c^\dagger_{y_n,\downarrow}\cdots c^\dagger_{y_1,\downarrow}\Omega)
\\
&=
\sum_{y_1<\cdots <y_n \leq 0 \leq x_1<\cdots < x_n} \prod_{j=1}^n (-1)^{x_j-y_j}
( c^\dagger_{x_n,\uparrow}\cdots c^\dagger_{x_1,\uparrow}
c^\dagger_{x_n,\downarrow}\cdots c^\dagger_{x_1,\downarrow}\Omega,
e^{-it \mathcal{H}}
c^\dagger_{y_n,\uparrow}\cdots c^\dagger_{y_1,\uparrow}
c^\dagger_{y_n,\downarrow}\cdots c^\dagger_{y_1,\downarrow}\Omega)
}
Here, we used the property $P\ket{0}=\ket{0}$ in the third line
and the transformation rule $U c_{x,\downarrow}U^\dagger=(-1)^x c_{x,\downarrow}$ in the fourth line.
We introduce the double-ket notation for operators and the Hilbert-Schmidt inner product:
\eq{
A \to \sket{A},\qquad (A,B) \to \sbra{A} B\rangle \! \rangle.
}
In this notation, superoperators are naturally regarded as operators acting on the corresponding double-ket states.
Note that the double-ket notation is introduced to avoid confusion with states in the original Hilbert space.
We then define
\eq{
\vert\Phi(\bm{x}) \rangle\!\rangle := \prod_{j=1}^n c^\dagger_{x_j,\uparrow} c^\dagger_{x_j,\downarrow} \sket{\Omega},
}
where $\Omega = \ket{0}\bra{0}$.
With this double-ket notation, the last expression can be rewritten as
\eq{
q_n(t)=\sum_{y_1<\cdots <y_n \leq 0 \leq x_1<\cdots < x_n} \prod_{j=1}^n (-1)^{x_j-y_j}
\sbra{ \Phi(\bm{x})} \vert e^{-it\mathcal{H}} \sket{ \Phi(\bm{y}) }.
}
Thus we obtain Eq.~\eqref{SM:map_hubbard}.

\section{Asymptotic analysis for the CGF}
\label{sec:asym}
We perform the asymptotic analysis of the CGF.
From Eq.~\eqref{eq:freddet} in the main text, the CGF is expressed as
\eqs{
\log \braket{e^{\lambda Q_t}} &= \log \det[\hat{I}+\omega \hat{K}^{(\gamma,t)}]_{[0,t]}
\\
&= \mathrm{Tr}\log [\hat{I}+\omega \hat{K}^{(\gamma,t)}]_{[0,t]}
\\
&= - \sum_{n> 0} \frac{(-\omega)^n}{n} \mathrm{Tr}[(\hat{K}^{(\gamma,t)})^n ]_{[0,t]},
}
where 
\eq{
\mathrm{Tr}[(\hat{K}^{(\gamma,t)})^n]_{[0,t]} = \prod_{j=1}^n \Big[\int_0^t du_j \Big]\;\prod_{j=1}^n K^{(\gamma,t)}(u_j,u_{j+1})
\label{eq:trace}
}
with the convention $u_{n+1} = u_1$.
Hereafter, we evaluate the asymptotic behavior of $\mathrm{Tr}[(\hat{K}^{(\gamma,t)})^n ]_{[0,t]}$ 
in the two limiting cases: the long-time limit $t\to \infty$ with finite $\gamma>0$ and the weak dephasing $\gamma \to 0$
and long-time $t\to \infty$ limit with fixed $\gamma t$.

\subsection{Diffusive current fluctuations in the long-time limit}
We consider the long-time limit $t\to\infty$ with the finite dephasing $\gamma>0$ for the CGF.

In order to get a suitable expression for the asymptotic analysis,
we make the substitutions $z\to \sqrt{u/(2t-u)}z$ and $w \to \sqrt{v/(2t-v)}w$ in Eq.~\eqref{eq:kernel} of the main text.
Then the kernel can be rewritten as
\eq{
K^{(\gamma,t)}(u,v) = e^{-2\gamma (u+v)} \oint \frac{dz}{2\pi iz} \oint \frac{dw}{2\pi i w} 
\frac{(1+ \frac{u}{2t-u} z^2)(1- \frac{v}{2t-v} w^2)
e^{\sqrt{u(2t-u)}(z-1/z)} e^{\sqrt{v(2t-v)}(1/w-w)}
}{(\sqrt{\frac{u}{2t-u}}z - \sqrt{\frac{v}{2t-v}}w )(1+\sqrt{\frac{uv}{(2t-u)(2t-v)}}zw) }. 
}
Here the integration contours can be taken to be the unit circles $|z|=|w|=1$ when $u\neq v$, 
while for $u=v$ they are slightly deformed so that the integrand is regular on the contours.
Note that the saddle points of $z-1/z$ are simply given by $z=\pm i$.
Hence,
we can perform the stationary phase method for large $t$ and finite $u,v$, obtaining the asymptotic form of the kernel
\eq{
K^{(\gamma,t)}(u,v) \simeq \frac{e^{-2\gamma (u+v)}  }{2\pi (uv)^{1/4}} \left(
\frac{\sin[2\sqrt{2t}(\sqrt{u}-\sqrt{v})]}{\sqrt{u}-\sqrt{v}} -\frac{\cos [2\sqrt{2t}(\sqrt{u}+\sqrt{v})]}{\sqrt{u}+\sqrt{v}}
\right).
\label{eq:kernel_long}
}

Substituting the above expression into Eq.~\eqref{eq:trace}, we obtain
\eqnn{
\mathrm{Tr}[(\hat{K}^{(\gamma,t)})^n ]_{[0,t]} 
&\simeq \prod_{j=1}^n \Big[\int_0^\infty \frac{du_j}{2\pi \sqrt{u_j} } e^{-4\gamma u_j} \Big] 
\prod_{j=1}^n 
\left(
\frac{\sin[2\sqrt{2t}(\sqrt{u_j}-\sqrt{u_{j+1}})]}{\sqrt{u_j}-\sqrt{u_{j+1}}} -\frac{\cos [2\sqrt{2t}(\sqrt{u_j}+\sqrt{u_{j+1}})]}{\sqrt{u_j}+\sqrt{u_{j+1}}}
\right),
\\
&= \prod_{j=1}^n \Big[\int_0^\infty \frac{du_j}{\pi } e^{-4\gamma u_j^2} \Big]
\prod_{j=1}^n 
\left(
\frac{\sin[2\sqrt{2t}(u_j-u_{j+1})]}{u_j-u_{j+1}} -\frac{\cos [2\sqrt{2t}(u_j+u_{j+1})]}{u_j+u_{j+1}}
\right),
}
where we extended the upper limit of the integral from $t$ to infinity in the first line and made the substitutions $u_j \to u_j^2$ in the second line.
Since the cosine term is of order one and rapidly oscillating, its contribution is suppressed. In contrast, the sine term becomes $\order{t^{1/2}}$ when
$|u_j-u_{j+1}| \leq \order{t^{-1/2}}$, Hence the leading contribution can be approximated as
\eq{
\mathrm{Tr}[(\hat{K}^{(\gamma,t)})^n ]_{[0,t]} \simeq 
\prod_{j=1}^n \Big[\int_0^\infty \frac{du_j}{\pi } e^{-4\gamma u_j^2} \Big]
\prod_{j=1}^n 
\frac{\sin[2\sqrt{2t}(u_j-u_{j+1})]}{u_j-u_{j+1}}.\label{eq:trace_long}
}
Since it is known that $\lim_{A\to \infty}\sin Ax/x = \pi \delta(x)$,
the above equation can be evaluated as
\eq{
\mathrm{Tr}[(\hat{K}^{(\gamma,t)})^n ]_{[0,t]} \simeq 2\sqrt{2t}\int_0^\infty \frac{du_1}{\pi} e^{-4n\gamma u_1^2} = \sqrt{\frac{t}{2n\gamma \pi}}.
}

Thus we obtain the asymptotic form of the cumulant generating function
\eq{
\log \braket{e^{\lambda Q_t}} \simeq -\sqrt{\frac{t}{2\gamma\pi}}\sum_{n\geq1} \frac{(-\omega)^{n}}{n^{3/2}} 
= \sqrt{\frac{t}{2\gamma}}\int_{-\infty}^{\infty}\frac{dk}{\pi} \log [1+\omega e^{-k^2}]. 
}

\subsection{Crossover from ballistic to diffusive behavior in the weak-dephasing regime}
We consider the simultaneous limits $t\to \infty$ and $\gamma \to 0$ with fixed $\gamma t$ for the cumulant generating function.
In this case, it is appropriate to consider the following rescaled kernel,
\eq{
t K^{(\gamma,t)}(tu,tv) = te^{-2\gamma t (u+v)} \oint \frac{dz}{2\pi iz} \oint \frac{dw}{2\pi i w} 
\frac{(1+ \frac{u}{2-u} z^2)(1- \frac{v}{2-v} w^2)
e^{t \sqrt{u(2-u)}(z-1/z)} e^{t \sqrt{v(2-v)}(1/w-w)}
}{(\sqrt{\frac{u}{2-u}}z - \sqrt{\frac{v}{2-v}}w )(1+\sqrt{\frac{uv}{(2-u)(2-v)}}zw) },
\label{eq:scaled_kernel}
}
since $\exp(-2\gamma t)$ is no longer small. 
Using this rescaled kernel, one can rewrite Eq.~\eqref{eq:trace} as
\eq{
\mathrm{Tr}[(K^{(\gamma,t)} )^n]_{[0,t]} = \prod_{j=1}^n \Big[\int_0^1 du_j\Big]\;  \prod_{j=1}^n
[tK^{(\gamma,t)}(tu_j,tu_{j+1})].
}

Similar to the derivation of Eq.~\eqref{eq:kernel_long}, the above equation for finite $u$ and $v$ can be approximated as
\eqnn{
 t K^{(\gamma,t)}(tu,tv)  \simeq 
 &-\frac{e^{-2\gamma t (u+v)} }{2\pi  [u(2-u)v(2-v)]^{\frac{1}{4}}} 
  \frac{1-\sqrt{\frac{uv}{(2-u)(2-v)}}}{\sqrt{\frac{u}{2-u}} + \sqrt{\frac{v}{2-v}}} \cos 2t[\sqrt{u(2-u)}+\sqrt{v(2-v)}]
  \\
  &+ \frac{e^{-2\gamma t (u+v)} }{2\pi  [u(2-u)v(2-v)]^{\frac{1}{4}}} \frac{\sqrt{\frac{u}{2-u}} -\sqrt{\frac{v}{2-v}}}{1+\sqrt{\frac{uv}{(2-u)(2-v)}}} 
   \cos 2t[\sqrt{u(2-u)}+\sqrt{v(2-v)}]
   \\
  & - \frac{e^{-2\gamma t (u+v)} }{2\pi  [u(2-u)v(2-v)]^{\frac{1}{4}}} \frac{\sqrt{\frac{u}{2-u}} +\sqrt{\frac{v}{2-v}}}{1-\sqrt{\frac{uv}{(2-u)(2-v)}}} 
   \sin 2t[\sqrt{u(2-u)}-\sqrt{v(2-v)}] 
   \\
   &
   + \frac{e^{-2\gamma t (u+v)} }{2\pi  [u(2-u)v(2-v)]^{\frac{1}{4}}} 
  \frac{1+\sqrt{\frac{uv}{(2-u)(2-v)}}}{\sqrt{\frac{u}{2-u}} - \sqrt{\frac{v}{2-v}}} \sin 2t[\sqrt{u(2-u)}-\sqrt{v(2-v)}].
}
Hence we obtain
\eqs{
\mathrm{Tr}[(\hat{K}^{(\gamma,t)})^n]_{[0,t]} &\simeq 
 \prod_{j=1}^n\Big[ \int_0^1 \frac{ds_j}{\pi} e^{-4\gamma t(1-\sqrt{1-s_j^2})}\Big]
 \prod_{j=1}^n \Big[\frac{\sin[2t(s_j - s_{j+1})]}{s_j - s_{j+1}} + \frac{\cos[2t(s_j + s_{j+1})]}{s_j + s_{j+1}}\Big]
 \\
 &\simeq  \prod_{j=1}^n\Big[ \int_0^1 \frac{ds_j}{\pi} e^{-4\gamma t(1-\sqrt{1-s_j^2})}\Big] 
  \prod_{j=1}^n \Big[\frac{\sin[2t(s_j - s_{j+1})]}{s_j - s_{j+1}} \Big]
}
where we changed the variables to $s_j := \sqrt{u_j(2-u_j)}$ in the first line and neglected the cosine terms in the second line.
In the above equation, we use the relation $\lim_{A\to \infty } \sin Ax/x = \pi \delta(x)$, obtaining
\eq{
\mathrm{Tr}[(\hat{K}^{(\gamma,t)})^n]_{[0,t]} \simeq \frac{2t}{\pi }\int_0^1 ds \; e^{-4n \gamma t(1-\sqrt{1-s^2})} .
}
Thus we obtain the asymptotic form of the cumulant generating function in the limits $\gamma\to 0$ and $t\to \infty$ at fixed $\gamma t $ as
\eq{
\log \braket{e^{\lambda Q_t}} \simeq  \frac{2t}{\pi }\int_0^1 ds \log (1+\omega \;e^{-4\gamma t (1-\sqrt{1-s^2})}).
}

\section{Numerical scheme for the CGF and the second cumulant}\label{sec:num}
We describe the numerical methods used to compute the CGF in the long-time limit ($t\gg 1$) 
for the domain wall initial condition ($\rho_a=1$, $\rho_b=0$), 
as well as the second cumulant in the regime $t\gg 1$ and $\gamma\ll 1$ for the steady state ($\rho_a=\rho_b=1/2$).
In our numerical simulations, we consider a finite lattice defined as $\Lambda_{L}=\{x\in \mathbb{Z}\mid -L<x\leq L\}$ with open boundary conditions.

\subsection{Numerical evaluation of the CGF for $\rho_a=1$ and $\rho_b=0$}
From Eq.~\eqref{eq:mom_genefun} in the main text, 
the CGF for $\rho_a=1$ and $\rho_b=0$ can be written as
\eq{
\log \braket{e^{\lambda Q_t}} = \log\Tr[e^{\lambda N_r}e^{\mathcal{L}t}[\ket{\mrm{DW}}\bra{\mrm{DW}}]  ],
\label{SM:cumufun_num}
}
where $\ket{DW}:= \prod_{x\leq 0} a^\dagger_x \ket{0}$. We numerically evaluate Eq.~\eqref{SM:cumufun_num}
using the unitary unraveling of the GKSL equation~\cite{Cao_2019,Eissler_unraveling}.

The unraveled dynamics are described by the following stochastic Schr\"{o}dinger equation,
\eq{
i d\vert \psi_t\rangle = H dt \vert \psi_t\rangle + \sum_{j} (\sqrt{4\gamma} n_j dW_t^j -2i\gamma n_i dt) \ket{\psi_t},
\label{unitary_unravelling}
}
where $dW^j_t$ represents the standard increment of the Wiener process with expectation values, 
$\mathbb{E}[dW^j_t]=0$ and $\mathbb{E}[dW_t^j dW_t^k]=\delta_{j,k}dt$,
and the multiplicative noise is understood in the It\^{o} convention.
Here, we denote the ensemble average over the Wiener process by $\mathbb{E}$.
Using the properties of the Wiener process, one can verify that the time evolution of $\mathbb{E}[\ket{\psi_t}\bra{\psi_t}]$
obeys the GKSL equation given in Eq.~\eqref{eq:GKSL} of the main text.

Since the generator of the time evolution for the unraveled state is quadratic, and the initial condition 
is a pure gaussian state $\ket{\mrm{DW}}$, one can apply Wick's theorem~\cite{GAUDIN196089} for each unraveled state.
In particular, we have the determinant representation of $\braket{e^{\lambda Q_t}}$ ~\cite{Schonhammer_2007},
\eq{
\braket{e^{\lambda Q_t}} =  \mathbb{E}[\det[\delta_{j,k}+(e^{\lambda}-1)\bra{\psi_t}a^\dagger_{j}a_k\ket{\psi_t}]_{j,k=1}^L].
}
Thus, the calculation reduces to evaluating two-point correlation function
$\bra{\psi_t}a^\dagger_{j}a_k\ket{\psi_t}$ for each realization.
To compute $\bra{\psi_t}a^\dagger_{j}a_k\ket{\psi_t}$ numerically, we employ the following numerical scheme~\cite{Cao_2019,Alberton_entanglement,Eissler_unraveling}.
Due to the quadratic nature of the time evolution, the unraveled state can be always written as
\eq{
\ket{\psi_t} = \prod_{k=-L+1}^0 \left[\sum_{j=-L+1}^L U_{j,k}(t)a^\dagger_j\right]\ket{0}
}
with the normalization condition $U^\dagger U=1$. 
Then it follows that $\bra{\psi_t}a^\dagger_ja_k\ket{\psi_t}=(UU^\dagger)^{*}_{j,k}$. 
Furthermore, from Eq.~\eqref{unitary_unravelling},
we have $U(t+dt)=e^{-i\sqrt{4\gamma}dW_t}e^{-ihdt}U(t)$
with $h_{j,k}:=-\delta_{j+1,k}(1-\delta_{k,-L+1})-\delta_{j-1,k}(1-\delta_{k,L})$
and $(dW_t)_{j,k}:=\delta_{j,k}dW_t^j$.
This formulation allows for efficient numerical simulations of the system dynamics.

\subsection{Numerical evaluation of the second cumulant for $\rho_a=\rho_b=1/2$}
To numerically investigate the crossover behavior of current fluctuations, we need to access the regime
satisfying the following conditions:
\eq{
t\gg1/\gamma \gg 1,
}
and 
\eq{
L \gg \sqrt{\frac{t}{2\gamma}}.
}
Although the numerical method presented in the previous subsection enables efficient computation, 
accessing this regime remains computationally demanding.
We therefore restrict our numerical analysis to the second cumulant $\braket{Q^2}_c$ 
in the steady state with $\rho_a=\rho_b=1/2$.
In this case, Eqs.~\eqref{eq:omega_dep} and \eqref{eq:q_n} yield
\eq{
\braket{Q^2}_c = \frac{1}{2} \sum_{y\leq 0 < x}\bra{x} e^{\mathcal{L}t}[\ket{y}\bra{y}]\ket{x}.
}
This quantity can be evaluated straightforwardly using the Runge--Kutta method, as it involves only a single-particle 
density-matrix element $\bra{x} e^{\mathcal{L}t}[\ket{y}\bra{y}]\ket{x}$.

\end{document}